\documentclass[lettersize,journal]{IEEEtran}
\usepackage{amsmath,amsfonts}
\usepackage{algorithmic}
\usepackage{algorithm}
\usepackage{array}
\usepackage[caption=false,font=normalsize,labelfont=sf,textfont=sf]{subfig}
\usepackage{textcomp}
\usepackage{stfloats}
\usepackage{url}
\usepackage{verbatim}
\usepackage{graphicx}
\usepackage{cite}
\begin{document}

\title{Nonlinear Stability Boundary Assessment Of Wind Power Plants Based on Reverse-Time Trajectory}

\author{
Sujay~Ghosh, \emph{Member IEEE},
	Mohammad Kazem Bakhshizadeh, Guangya Yang, \L{}ukasz Kocewiak

\thanks{		
		S. Ghosh, M. K. Bakhshizadeh and \L{}. Kocewiak are with \O{}rsted Wind Power, Nesa Allé 1, 2820, Denmark, (e-mail: sujgh@orsted.com). 
		
		G. Yang is with the Technical University of Denmark, Anker Engelunds Vej 1, 2800, Denmark.}
\thanks{Manuscript received April 19, 2021; revised August 16, 2021.}}

\markboth{}
{S. Ghosh \MakeLowercase{\textit{et al.}}: Nonlinear Stability Assessment Of Wind Power Plants Based on Reverse Time Trajectory}

\maketitle

\begin{abstract}
This letter determines the nonlinear stability boundary of a wind power plant (WPP) connected to an AC power grid via a long HVAC cable. The analysis focuses on the slow Phase-Locked Loop (PLL) dynamics, with an assumption that the fast current control dynamics can be neglected. To begin, we propose an aggregated reduced-order wind turbine model. This aggregation can be applied up to a limited frequency, e.g. 400Hz, which aligns with our assumption regarding low-frequency dynamics. The WPP collector and transmission network model is established using impedance/frequency scan approximated around $\pm$5 Hz of the PLL nominal frequency, accounting for the hard saturation limits. The stability boundary of the reduced-order system is determined by reverse time trajectory, offering valuable insights into the WPP's overall stability. The work presents a routine from modelling to nonlinear stability assessment for offshore wind farm applications.
\end{abstract}


\section{Introduction}
\IEEEPARstart{T}{he} installation of wind power plants (WPPs) has notably increased globally to meet the rising demand for sustainable energy sources. However, the WPPs' remote location, high power output relative to the grid's strength, and potential to exhibit rapid and complex transients during grid disturbances have necessitated the assessment of system stability concerns. Existing literature has presented excellent results on the transient stability of power converters \cite{ref1}-\cite{ref3}, and identified the Phase Locked Loop (PLL) as a key contributor to instabilities driven by large-signal disturbances. 

While many studies have examined stability boundaries by modelling single WT units in simplified AC grid contexts \cite{ref1}-\cite{ref3}, few have explored extending this assessment to the WPP level. However, these attempts often involve a single machine equivalent without considering the influence of aggregation on boundary estimation. Furthermore, some studies simplify collection lines to a degree that may not accurately capture real-world complexities. This work aims to bridge this research gap by developing a methodology that addresses these limitations.

The main contribution of this letter is to present a modelling method for nonlinear stability analysis of a WPP interconnection. Thereafter, the stability boundary of the resulting reduced-order system is determined by the reverse time trajectory, offering valuable insights into the WPP's overall stability. 

\begin{figure}[!t]
\centering
\includegraphics[width=9.25cm]{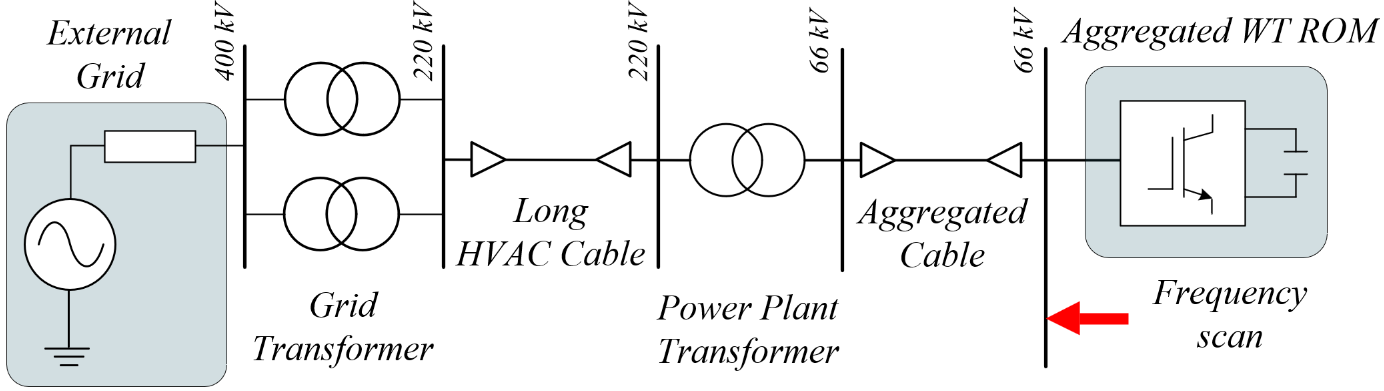}
\caption{WPP with aggregated WTs connected to an AC power grid.}
\label{fig_1}
\end{figure}

\section{Problem Formulation}
\subsection{Concepts of Nonlinear Stability}
A nonlinear dynamical system may be mathematically represented by a set of differential equations (ODEs) as follows:

\begin{equation}\label{eq:1}
      \dot{x} = f(t, x, u)\\
\end{equation}
where $t$ denotes time, $\dot{\textbf{\textit{x}}}$ represents the time derivative of the state vector $\textbf{\textit{x}}\in\mathbb{R}^n$, and $\textbf{\textit{u}}\in\mathbb{R}^m$ is the input vector.
If the system (1) possesses a stable post-disturbance equilibrium $x_0$, characterised by $f$($x_0$) = 0, then the main problem of nonlinear stability lies in determining whether the system can reach $x_0$ from the post-disturbance condition $x_f$.

The region of attraction (RoA) comprises all the $x_f$ points from which trajectories converge to the equilibrium point $x_0$. Mathematically, this region can be defined as:

\begin{equation}
A(x_0) = {x| \lim_{t \to \infty} \Phi(t, x) = x_0 }
\end{equation}

Thus, the primary objectives of nonlinear stability are:
\begin{itemize}
\item  Identifying the post-disturbance condition $x_f$, as it depends upon the type of disturbance.
\item  Evaluating whether $x_f$ falls within the estimated RoA.
\end{itemize}

\subsection{Modelling of Wind power plant}
As indicated in Fig. 1, this paper considers a WPP with aggregated WTs, where the transmission network is represented as a long HVAC cable connected to a simple grid equivalent. To analyse the nonlinear stability of the WTs, the state variables of the inner current control are reduced to their steady-state values since the current control dynamics are much faster than that of the PLL. 
\begin{figure}[h]
    \centering
    \includegraphics[width=9.0cm]{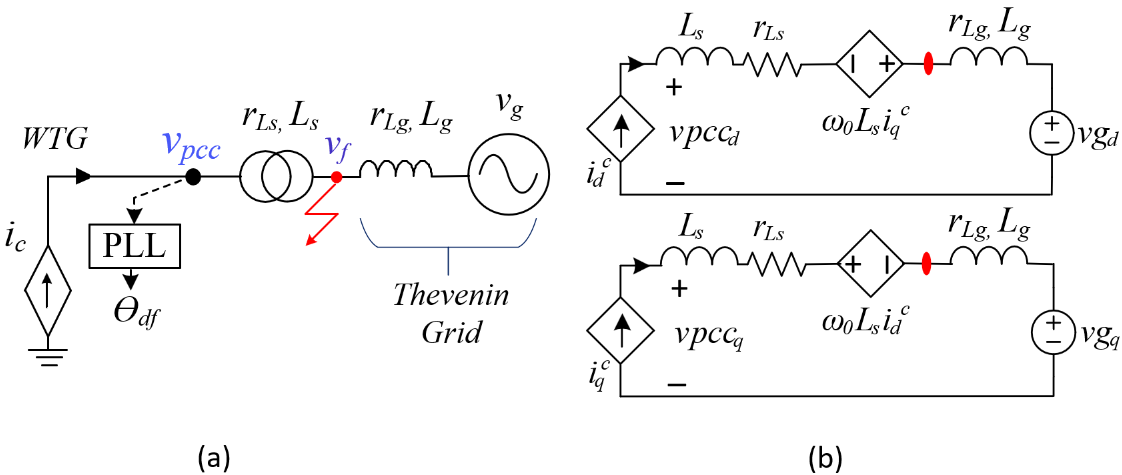}
    \caption{Wind turbine model: (a) Reduced order model (ROM) of the Type-4 wind turbine considering the actions/assumptions. (b) System representation of ROM in the DQ domain.}
    \label{ROM}
\end{figure}
Figure 2 presents the reduced-order representation of the WPP, where $v_{pcc}$ is the voltage at point of common coupling; $r_{Ls}$ and $L_s$ are the WT transformer resistance and inductance, respectively; $i_c$ is the current injected by the WTs; and $V_g$, $r_{Lg}$ and $L_g$ are the Thevinin equivalent of the WPP transmission network. It's important to note that in this study, aggregated WTs are a viable consideration, as their accuracy holds within a defined frequency range, typically up to 400Hz, as supported by reference \cite{ref5}. This aligns seamlessly with our underlying assumption regarding the relevance of low-frequency dynamics.
The reduced order model (ROM) of a single WT unit \cite{ref6} can be presented as,

\begin{equation}\label{SEq_1}
\begin{aligned}
\dot{x_1} &= x_2 \\
x_{2} &= x_{2}^{max} \cdot \text{tanh}( {x_{3}}/{x_{2}^{max}})\\
M_{eq} \dot{x_3} &= T_{m_{eq}} - T_{e_{eq}} - D_{eq} {x_2}\\
\end{aligned}
\end{equation}

where, $x_1= \delta$, $x_2= \dot{\delta}$ (saturated) and $x_3= \dot{\delta}$,
\begin{equation}\label{SEq_2}
\begin{aligned}
M_{eq} &= 1- k_p L_g i_d^c\\
T_{m_{eq}} 
&= k_p( \dot{\overline{r_{Lg} i_q^c}} + \ddot{\overline{L_g i_q^c}} + \dot{\overline{L_g i_d^c}} \omega_g)
+ k_i( r_{L_g} i_q^c \\
&\qquad+ \dot{\overline{L_g i_q^c}} + L_g i_d^c \omega_g) \\
T_{e_{eq}} 
&=  (k_i V_f \text{sin} x_1 + k_p \dot{V_f} \text{sin} x_1) + M_{eq} \dot{\omega}_g\\
D_{eq} 
&= k_p ( V_f \text{cos}x_1 - \dot{\overline{L_g i_d^c}}) - k_i L_g i_d^c
\end{aligned}
\end{equation}

The system (3) is modelled in a DQ frame rotating at a fixed frequency $\omega_0$. 
The mathematical model for the WT system and its controls is discussed in detail in \cite{ref6}. During faults at WT terminal, the voltage $V_f$ and the current injection $i_d^c$ and $i_q^c$ can be computed as,
\begin{equation}\label{}
\begin{aligned}
V_{f,3ph} &= \frac{Z_{f}}{Z_{g} + Z_f} \cdot \left[ vg_{dq} +  i_c \cdot  Z_{g} \right]\\
i_{q,3ph}^c &= K_{factor} \cdot v_{pcc} \\
i_{d,3ph}^c &= \sqrt{I_{max}^2 - (i_{q,3ph}^c)^2} \\
\end{aligned}
\end{equation}
where, $Z_f$ is the fault impedance, $Z_{g}= (r_{Lg} +j\cdot \omega_0 L_g$), and $i_{q,3ph}^c$ is limited to 1 pu.
Consequently, based on aggregation characteristics, considering $N$ turbines in the WPP, the system (3) can be scaled up as, 
\begin{equation}\label{}
\begin{aligned}
i_{d,eq}^c &= (N \cdot i_d^c),\quad i_{q,eq}^c = (N \cdot i_q^c)\\
r_{Ls,eq} &= (r_{Ls} / N),\quad L_{s,eq} = (L_{s} / N) 
\end{aligned}
\end{equation}
Furthermore, the impedance of the WPP transmission network $r_{Lg}$ and $L_{g}$ can be obtained by a frequency scan, and curve fitted around $\pm$5 Hz of the PLL frequency due to the hard saturation limits, such that,

\begin{equation}\label{}
\begin{aligned}
r_{Lg} &= \mathbb{R}\{Z(F_{C})\} \\
L_{g} &= m/ (2\pi) \\
\end{aligned}
\end{equation}
where, $Z(f)$ is the impedance-frequency scan vector, $F_C$ is the corner frequency of the frequency scan, and $m$ is the slope of the frequency scan around the $\pm$5 Hz of the PLL nominal frequency. For more information, refer to Section IV.

\section{Nonlinear Stability Analysis }
If the function $f$ in (1) satisfies the Lipschitz condition, it guarantees a unique trajectory for each initial condition. Moreover, solving the differential equations backwards in time results in traversing the same unique trajectory. This is called 'Reverse-time trajectory'. It implies that (1) can be solved as (8) with initial conditions close to the equilibrium point.

\begin{equation}\label{}
    \dot{x} = -f(t, x, u)\\
\end{equation}
The response of the dynamical system (8) after a given time $t$, with initial conditions chosen from a closed set $\mathbb{B} \subset \mathbb{R}^n$, will lie inside a closed set $\mathbb{D} \subset \mathbb{R}^n$ \cite{ref7}; i.e. for any initial point inside a boundary  $\mathbb{B}$, the final response lies in the calculated final boundary $\mathbb{D}$, which we call as a 'time-limited region of attraction' (TLRoA) denoted by $A_{TL}(t, x_0)$.

A TLRoA is a subset of the actual system RoA $A(x_0)$, i.e. any point at its boundary will attract to the equilibrium at the selected time $t$, and any point inside will attract to the equilibrium in time less than $t$. 
The initial points for $A_{TL}(t, x_0)$ can be computed from the boundary of a linearised Lyapunov function $V(\textbf{\textit{x}}) = \textbf{\textit{x}}^T\textbf{P}\textbf{\textit{x}} $ as discussed in \cite{ref7}. A sample TLRoA with different time limits is illustrated in Fig. 3 and is compared against its forward simulated RoA.   

\begin{figure}[h]
    \centering
    \includegraphics[width=7.0cm]{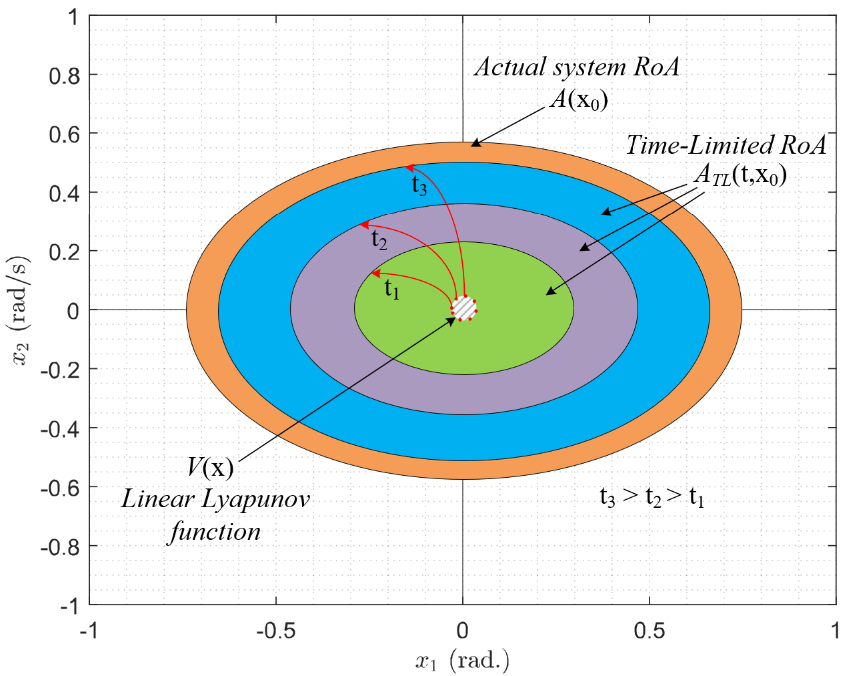}
    \caption{Comparison of RoA against TLRoA with $t_1$, $t_2$ and $t_3$, such that $t_3>t_2>t_1$.}
    \label{fig:Ramp1}
\end{figure}

\section{Case Study }
In this study, we investigate the effectiveness and accuracy of our proposed stability analysis by applying it to the test power system, as illustrated in Fig. 1. The system comprises a 180 MW WPP, achieved through aggregated three strings of five 12 MW WT units per string. To assess nonlinear stability, we consider various system parameters outlined in Table I.
 
\begin{table}[h]
\caption{SYSTEM AND CONTROL PARAMETERS}
\label{table}
\setlength{\tabcolsep}{3pt}
\begin{tabular}{|p{40pt}|p{105pt}|p{70pt}|}
\hline
Symbol& 
Description& 
Value \\
\hline
$S_b $& 
Rated power& 
3x5x12 MVA \\
$V_g$& 
Nominal converter voltage & 
690 $\sqrt{2/3}$ V\\
$\omega_0$& 
System angular speed & 
314 rad/s \\
$r_{Lg}$, $L_{g}$& 
WT transformer impedance & 
16.6 $\mu$ohms, 1.23 $\mu$H \\
$K_{cc}$& 
Fast current control: $k_p$ $I_i$ & 
0.5, 0.03 \\
$K_{pll}$& 
SRF PLL design: $k_p$ $k_i$ & 
0.025, 1.5 \\
$K_{factor}$& 
K-factor during LVRT  & 
2 \\
\hline
\end{tabular}
\label{tab1}
\end{table}

The impedance of the WPP transmission network, $r_{Lg}$ and $L_{g}$, is obtained by a frequency scan from the node highlighted in Fig. 1 and is approximated around $\pm$5 Hz of the PLL nominal frequency due to the hard saturation limits, as shown in Fig. 4. The slope ($m$) line which intersects the frequency-axis gives the corner frequency ($F_C$). Equation (7), referred to the low voltage side of the WT transformer, gives $r_{Lg}=98.5$ $\mu$ohm and $L_g=2.17$ $\mu$H.

\begin{figure}[h]
    \centering
    \includegraphics[width=6.250cm]{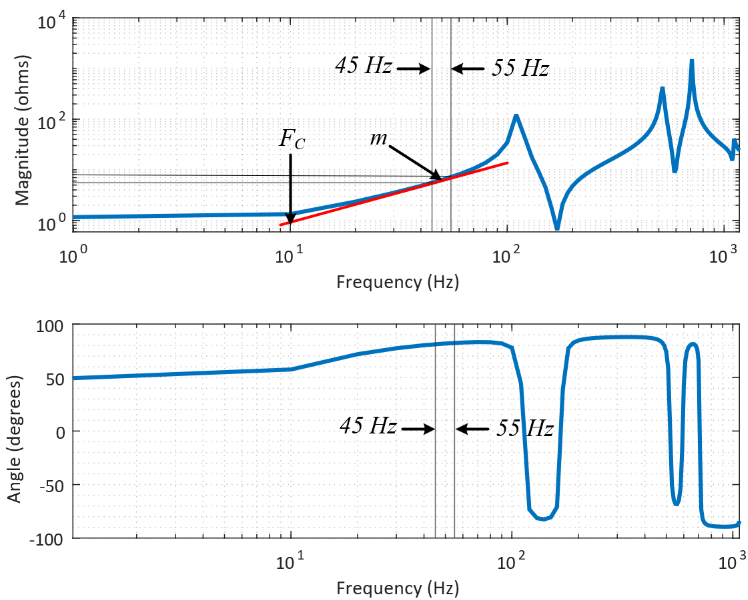}
    \caption{Impedance frequency scan of the WPP transmission network.}
    \label{fig:Ramp1}
\end{figure}

A balanced bolted fault is applied at the aggregated WT terminal at t = 1.5 s, recovering after 0.1 s with an active current ramp of 2 pu/s. Figure 5 shows a comparison of the system trajectories, validating the accuracy of the proposed WPP ROM against the actual EMT model in PSCAD. Overall, the proposed ROM aligns well with PSCAD simulations, except for high-frequency dynamics, which can be ignored when analysing the slow PLL dynamics. 

\begin{figure}[h]
    \centering
    \includegraphics[width=6.50cm]{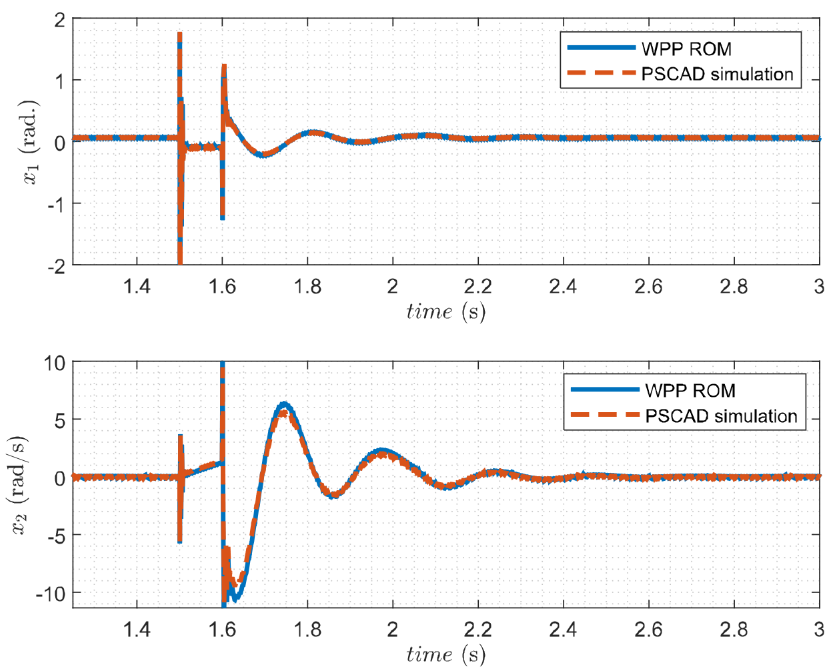}
    \caption{A comparison of the system trajectories, PSCAD vs ROM.}
    \label{fig:Ramp1}
\end{figure}

Figure 6 illustrates the stability boundaries (TLRoAs) of the WPP for two distinct cases: case-1 representing normal operation, and case-2 depicting a weaker grid scenario. The ROM (3)-(7) exhibits multiple equilibrium points that repeat every $\pm 2\pi$ rads, resulting in the WPP having multiple RoAs; more details can be found in \cite{ref7}. The TLRoAs are obtained with a time of 2.25 s. Notably, under a weaker grid, the size of the TLRoA reduces, indicating lower stability compared to case-1. The critical clearing time, i.e. the time for the fault trajectory to reach the TLRoA boundary, is found to be 0.89 s for case-1 and 0.84 s for case-2. 

\begin{figure}[h]
    \centering
    \includegraphics[width=6.50cm]{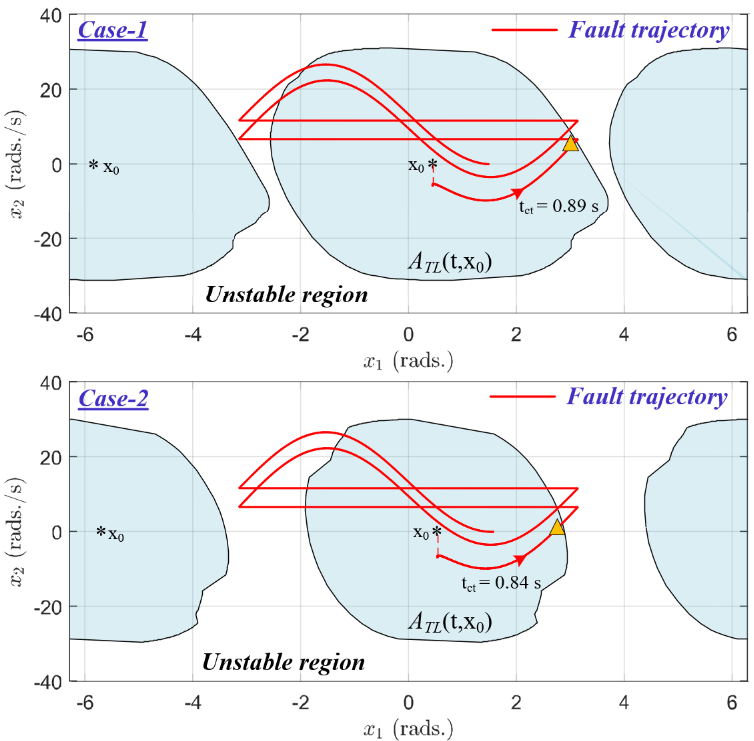}
    \caption{TLRoA of the test WPP; case-1 representing normal operation, and case-2 depicting the N-1 contingency of the grid transformer.}
    \label{fig:Ramp1}
\end{figure}

\section{Conclusion}
The main contribution of this letter is the proposal of a stability assessment routine. The derived stability boundary, i.e. the time-limited region of attraction (TLRoA), which is notably a subset of the actual system RoA, is obtained by reverse-time trajectory. 
The proposed method overcomes analytical limitations, enabling WPP scaling with nonlinear components like saturation and ramp rate limits, and demonstrates its applicability in stability analysis under large disturbances.
The studies highlight the importance of grid strength and demonstrate that the WPP's stability is compromised during grid contingencies, leading to a reduced size of TLRoA and potentially affecting the overall power system's reliability.


 




\vfill

\end{document}